\documentclass[12pt]{article}
\usepackage{amsmath,amsthm,amssymb,amsxtra, graphicx}
\usepackage{color}
\usepackage{amsmath, amsthm, amssymb}
\usepackage{graphicx}
\usepackage{dcolumn}
\usepackage{bm}
\usepackage{color}
\usepackage{subfigure}
\usepackage[latin1]{inputenc}
\newtheoremstyle{theorem}
{10pt} 
{10pt} 
{\sl} 
{\parindent} 
{\bf} 
{. } 
{ } 
{} 
\theoremstyle{theorem}
\newtheorem{theorem}{Theorem}

\newtheoremstyle{defi}
{10pt} 
{10pt} 
{\rm} 
{\parindent} 
{\bf} 
{. } 
{ } 
{} 
\theoremstyle{defi}

\begin{document}
	
	\pagestyle{empty}

	\title{\bf{Hyperbolic quantum color codes}}
	\author{WALDIR SILVA SOARES JR.$^1$, EDUARDO BRANDANI DA SILVA$^2$ \\
	\small $^{1}$Department of Mathematics - UTFPR \\
    \small 85503-390, Pato Branco - PR, Brazil \\\
	\small  DMA - Maringa State University - UEM$^2$  \\
	\small  Av. Colombo 5790 - Maring\'a - PR - 87020-000 - Brazil \\
	\small $^1$ dir.soares.junior@gmail.com\\
	\small $^2$ebsilva@uem.br}
	
	\date{}
	\maketitle

\begin{abstract}
Current work presents a new approach to quantum color codes on compact surfaces with genus $g \geq 2$ using the identification of these surfaces with hyperbolic polygons and  hyperbolic tessellations. We show that this method may give rise to color codes with a very good parameters and we present tables with several examples of these codes whose parameters had not been shown before. We also present a family of codes with minimum distance $d=4$ and the encoding rate asymptotically going to 1 while $n \rightarrow \infty$. \\

{\bf keywords:} quantum color codes, quantum error-correcting codes, surface codes, tessellations, hyperbolic geometry
\end{abstract}

\section{Introduction}           

\noindent The concept of error-correcting codes (classical) was introduced by Claude Shannon in 1948, and since his seminal works, the theory has been developed in great depth and become  indispensable to the development of Information and Communication theories.

The ideas of the classical error-correction theory were inspiration and became models for the creation of 
quantum error-correcting codes. In 1996 there was a breakthrough in quantum coding by the discovery of a class of codes, now known as CSS codes, by Robert Calderbank, Peter Shor and Andrew Steane, as can be seen in \cite{Steane1996Simple}, \cite{CSSexists}, and which generated a rich code structure, which is the Stabilizer Quantum Codes \cite{Gottesman1996Class}.

Kitaev, \cite{Kitaev}, proposed a particular case of the Stabilizer Quantum Codes, which are associated with a $\bf{Z}_2$ lattice. Since such codes depend on the topology of a surface, they were called Topological Codes. In topological quantum codes we encode the quantum words in the non-local degrees of freedom of strongly correlated quantum systems that have a topological order. Due to this non-local coding, these quantum words are intrinsically resistant to the disturbing effects of noise, as long as that noise is local in the sense of not affecting global topological properties of the system. This construction is based on an intrinsically physical mechanism that makes the topological system capable of self-correcting local errors.

Non-locality, which is a gain in the robustness of the system, generates some loss in another area of coding. In this case, as shown in \cite{bravyinogo}, in a family of $2D$ local codes, the distance is $ d = O(\sqrt{n})$, where $n$ is the number of qubits.

In \cite{Brandanigenus2} was proposed an extension of the Kitaev codes for surfaces with genus larger than one, using tools of hyperbolic geometry. Increasing the genus of the surface, Albuquerque, Palazzo and Silva used tessellations of the hyperbolic plane to create surface codes with the best parameters in the literature, until that moment. In these codes, some families generated by auto-dual, quasi auto-duals, and denser tessellations were highlighted, as later published in \cite{brandaniselfdual}.

After these, new results using hyperbolic geometry in the creation of better families of topological codes were obtained, and the attempt to a better understanding of the existing ones was made. 
In \cite{barbara}, the authors took an approach similar to that of \cite{Brandanigenus2}.  They give numerical estimates for the noise threshold and logical error probability of these codes against independent $X$ or $Z$ noise, assuming a noise-free error correction.

Bombin and Martin-Delgado, in \cite{distillation}, proposed a subclass of the stabilizer quantum codes, which became known as color quantum codes. The color quantum codes are also topological codes, like those of Kitaev's, but with an extra element in the labelling, the color. An immediate advantage of color coding is the fact that they encode twice as many qubits as surface codes, relative to the same surface.

Based on closed two-dimensional surfaces, in \cite{delfosse} is proved that 
\[
k d^2 \leq c (\log k )^2 n 
\]
where $c$ is a constant, it is valid for the homological codes, such as Kitaev's, surface and color codes. 

Regarding the color codes, a great step has already been taken for its real implementation, in the way to the construction of the quantum computer. In 2014, the authors of \cite{delgado}, using a special case of the color codes (triangular codes), implemented a quantum error correction code, encoding one qubit in entangled states distributed over $7$ trapped-ion qubits. This construction shown the capability of the code to detect one bit flip, phase flip or a combination of both errors, regardless of which of the qubits they occur.

The objective of this work is to make an approach inspired by \cite{Brandanigenus2}, to the color quantum codes environment. Analyzing the classes of codes generated by this technique, we obtain new families of codes with good properties. To achieve this goal we will work with compact surfaces of genus greater than or equal to two. These surfaces may be generated by a hyperbolic polygon that, via pairing of the edges, can be identified with such surface. These are the fundamental polygons for the surfaces. Among the infinite regular tessellations of the hyperbolic plane, we will calculate the possible subtessellations for each fundamental polygon that can generate color codes, analyzing their parameters and their codification rates.

We present some families of codes which are generated in this way, and we also show some codes with optimal parameters. Finally, we show that, for a given  family, the coding rate $\frac{k}{n} \rightarrow 1$ when $n \rightarrow \infty$, which is an important result.

Current work is organized as follows: Section $2$ presents a quick review of concepts of Hyperbolic Geometry, which are necessary to build the tessellations and to justify the relation between surfaces with genus greater than or equal to $2$, with fundamental hyperbolic polygons. In Section $3$ we give the concepts of stabilizer codes and their particular cases, which are Kitaev's toric codes and the Surface Codes. In Section $4$ we give the basic concepts of the quantum color codes. In Section 5, we have the main core of the work, where we apply the techniques and tools of Hyperbolic Geometry to generate families of color quantum codes on compact surfaces of genus greater than or equal to $2$. 

\section{A Glimpse for Hyperbolic Geometry}

\noindent Let us briefly go through some concepts of Hyperbolic Geometry that will enable us to make the necessary constructions for the development of this work. In case the reader wants to read something more detailed, we suggest references \cite{katok1992fuchsian}, \cite{stillwell1995geometry} and \cite{beardon2012geometry}.

A hyperbolic polygon $P$ with $p$ sides, called $p$-gon, is a closed and connected set formed by the region bounded by $p$ hyperbolic geodesic segments. The intersection of two adjacent geodesic segments is called a vertex. A $p$-gon in which all edges have the same length and the measure of their inner angles are all equal is called a regular $p$-gon.

If $A$ is a region of the hyperbolic plane $\bf{H}^2$, the hyperbolic area of $A$ is given by:
\[
\mu(A) = \int\int_A \frac{dx dy}{y^2} \,,
\]

\noindent if the integral exists.

However, to determine the area of a hyperbolic triangle, and hence any regular hyperbolic polygon, we can use the Gauss-Bonnet theorem, which shows that the area of a hyperbolic triangle depends only on its angles \cite{stillwell1995geometry}:

\begin{theorem} \label{teo GB} (Gauss-Bonnet) 
	Let $\Delta$ be a hyperbolic triangle with internal angles $\alpha,\beta,\theta$. Then, the area of $\Delta$ is given by 
	\[
	\mu(\Delta) = \pi - \alpha - \beta - \theta.
	\]
\end{theorem}

We denote by $PSL(2,\bf{R})$ the multiplicative group of the M\"obius transformations $T : \bf{C} \rightarrow \bf{C}$, which are defined by $T(z) = \frac{az + b}{cz + d}$, where $a, b, c, d \in \bf{R}$ satisfy $ad - bc = 1$. A Fuchsian Group is a discrete subgroup of $PSL(2,\bf{R})$. It is important to observe that all M\"obius transformations preserve the hyperbolic distance in $\bf{H}^2$, then they map geodesics to geodesics and also they are conformal transformations, that is, they preserve angles. It follows that the hyperbolic area of a hyperbolic polygon is invariant under the transformations of $PSL(2,\bf{R})$.

As may be seen in \cite{stillwell1995geometry}, any compact topological surface $\bf{M}$ may be obtained from a polygon $P^\prime$ by pairs of identified edges, since the side and angle conditions is satisfied.

The identification of the sides of a hyperbolic polygon $ P^{\prime}$ is defined as a side pairing transformation. Since the lengths of the sides are equal, a side pairing transformation is an isometry $\gamma \neq Id$ of an isometry group $\Gamma$, that preserves the orientation, taking a side $s$ of $P^\prime$ to another side $\gamma(s) = s^\prime$ of $P^\prime$, so that $\gamma^{-1} \in \Gamma \backslash \{Id\}$ and that $\gamma^{-1}(s^\prime)=s$. If $s$ is identified with $s^\prime$ and $s^\prime$ is identified with $s^{\prime \prime}$, then $s$ is identified with $s^{\prime \prime}$. This chain of identifications can also occur with vertices, and then we call a maximal set $\{v_1, v_2, ..., v_k \}$ of identified vertices of a vertex cycle \cite{Brandanigenus2}.

A pairing of the sides of $P^\prime$ defines an identification space $S_{P^\prime}$ where there is a distance function that coincides with the hyperbolic distance, for sufficiently small regions inside $P^\prime$, making it a hyperbolic surface when the sum of the angles of each vertex cycle is $2\pi$.

A compact surface $\bf{H}^2/\Gamma$ is the identification space of a polygon if the polygon is a fundamental region for $\Gamma$. A necessary and sufficient condition for a polygon to be a fundamental region is as follows \cite{stillwell1995geometry}:

\begin{theorem} \label{cond l.a.} (Side and Angle Conditions) If a compact polygon $P^\prime$ is a fundamental region for a group of isometries $\Gamma$ that preserves orientation in $\bf{S}^2$, $\bf{R}^2$ or $\bf{H}^2$, then
	
	\begin{itemize}
		\item for each side $s$ of $P^\prime$ there is exactly one other side $s^\prime$ of $P^\prime$ of the form $s^\prime = \gamma(s)$, for $\gamma \in \Gamma$ (the elements $\gamma$ are called side-pairing transformations
		of $P^\prime$);
		\item given a pairing of sides of $P^\prime$, for each set of identified vertices, the sum of the angles must be equal to $2\pi$. This set is a cycle of vertices. 
	\end{itemize}
\end{theorem}	

\begin{theorem} \label{poinc} (Poincar\'e)
	A compact polygon $P^\prime$ satisfying the side and angle conditions is a fundamental region for the group $\Gamma$ generated by the side pairing transformations of $P^\prime$, and $\Gamma$ is a Fuchsian group.
\end{theorem}

\section{Surface Stabilizer Codes}

A quantum error-correcting code (QECC) is an application of a complex Hilbert space $\mathcal{H}^k$, of dimension $2^k$, to a Hilbert space of dimension $2^n$ where $k < n$. A QECC $\mathcal{C}$ with code word length $n$, dimension $ k $ and minimum distance $ d $ is denoted by $[[n, k, d]]$.

The first classes of error-correcting quantum codes are those that dealt with bit-flip or phase-flip errors, or even a combination of them, such as Shor's Code \cite{shor1995}, proposed in 1995, which is a code based on the concatenation of a $3$-qubit code, that protects against bit-flip errors, with a $3$-qubits code that protects against phase-flip errors. Soon after, Steane proposed his $7$-qubits code \cite{Steane1996a}, which was not based on concatenations and it was shorter than the Shor's Code. Next, a class of codes, which comes from classical error-correcting linear codes, was introduced and became known as CSS codes (Calderbank, Shor, and Steane as can be seen in \cite{CSSexists}, \cite{Steane1996Simple}).

A more general class of codes, that even includes the CSS-like codes, are the  so-called Stabilizer Codes \cite{lidar2013quantum}. Stabilizer codes are the quantum analog for the classical additive codes.

To define these codes we consider the set formed by the Pauli matrices of a qubit $P = \{I, X, Y, Z \}$. We call a Pauli group of order $ n $ the set $P^n$ given by elements of the type $i^k P_1 \otimes P_2 \otimes \dots \otimes P_n$, where for every $i = 1 \ldots n$ one has $P_i \in P$ and $ k \in \{0,1,2,3\}$. Now consider an Abelian subgroup $S \subset P^n$ so that $ -I \notin S$, which we call the stabilizer group.

Thus, the stabilizer code $\mathcal{C}$ is defined by the auto space associated with the eigenvalue $1$ of the operators of $S$, that is: \cite{Gottesman1996Class} 
\[
\mathcal{C} = \{| \psi \rangle ; M | \psi \rangle = | \psi \rangle, \forall M \in S \} \,.
\]

If a stabilizer code has $r$ generators, then the dimension of the stabilizer subspace is $2^{n-r}$ \cite{lidar2013quantum}, implying that the number of coded qubits is:

\begin{equation} \label{codif}
k = n - r \,.
\end{equation}

In \cite{Kitaev}, Kitaev proposed a particular case of stabilizer code, which became known as Kitaev's Toric code. In a flat torus $l \times l$, Kitaev considered a square lattice tessellating the torus by regular polygons. He associated a qubit to each edge of the tessellation. At each vertex $v$ of the tessellation was associated an operator $X_v$ that acts as the matrix of Pauli $X$ on each edge adjacent to this vertex and as the identity in all others. Each face $f$ of the tessellation was associated with an operator $Z_f$ which acts as the matrix of Pauli $Z$ on each edge of the boundary of $f$, and act as identity in all others faces.

\begin{figure}[!h]
	\centering
	\includegraphics{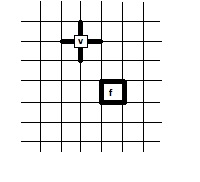}
	\caption{Support of face (f) e vertices (v) operators of Kitaev's Toric Code}
	\label{suportekitaev}
\end{figure}

In this way, Kitaev's Toric code is defined by:

\[
\mathcal{C}=\{|\psi \rangle; X_v|\psi\rangle=|\psi\rangle; Z_f|\psi\rangle=|\psi\rangle; \forall v, f \} \,.
\]

This code has parameters $[[2l^2,2,l]]$ where $k = 2$ is the number of non-trivial cycles, and $d = l$ is the number of edges contained in the shortest non-trivial homological cycle.

After, Kitaev's construction was generalized, given origin to the Surface Codes. For instance, surface codes were constructed using other subtessellations of the torus \cite{Bombin2006Topological}, which are generated by Lee spheres. In \cite{Brandani2009toric} was considered subtessellations constituted by polyominoes algebrically generated. Another extension of the toric codes was obtained using other surfaces with genus greater than one \cite{Brandanigenus2}, thus increasing the number of coded qubits without the need to add boudaries or "holes".

\section{Quantum Color Codes}

The Quantum Color Codes were introduced by Bombim and Martin-Delgado in \cite{distillation}.

To generate a color code on a two-dimensional surface, we need a trivalent, $3$-colorable tessellation. Trivalent tessellation means that each vertex of the tessellation is the intersection of exactly 3 edges, and to say that this tessellation is $3$-colorable means that it is possible to color all the faces of it using only $3$ colors (for instance, red, green and blue) so that two faces that share a common edge do not have the same color. With a coloration of the faces in this way, we may also induce a coloration of the edges, so that an edge of a certain color does not belong to the border of a face with that same color.

Unlike the surface codes, in the color codes the qubits are associated to the vertices of the tessellation and the generators of the stabilizers are the face operators (plaquette operators), and there are both $X$ and $Z$ operators in each face which act in the vertices of the face. For each face $p$ of the tessellation, we denote such operators by $B^{\sigma}_p$ with $\sigma = X, Z$.

Separating the faces according to their color, in the sets $R$ (red), $G$ (green) and $B$ (blue), one has:

\begin{equation} \label{color face}
\prod_{p\in R} B^{\sigma}_p = \prod_{p\in G} B^{\sigma}_p = \prod_{p\in B} B^{\sigma}_p
\end{equation}
\noindent with $\sigma = X, Z$.

Now we introduce the concept of {\itshape shrunk lattice}, one for each color, which are auxiliary lattices. In the red shrunk lattice, for instance, on each red face we place a vertex. Each edge of the shrunk lattice corresponds to two vertices of the original tessellation, and the green and blue faces of the colored tessellation are the faces of this auxiliary lattice.

The equation (\ref{color face}) implies that four of these generators are not independent, and as we may see in (\ref{codif}), this allows us to calculate the number of coded qubits. Now we can apply this to a shrunk lattice, with a color fixed: since the number of generators is $r = 2(V + F-2)$ and the number of physical qubits is twice the number of edges $(n = 2E)$ one has \cite{Espanhol}

\begin{equation}
k = n-r = 4 - 2 \mathcal{X}
\end{equation}

\noindent where $\mathcal{X}$ is the Euler characteristic of the surface. Note that the number $k$ of coded qubits depends only on the surface considered, not the tessellation used.

String operators are fundamental for the color codes. As in the surface codes, the homology of these strings is defined on $\bf{Z}_2$, since we are dealing with two-level quantum systems.

These strings may be green, blue or red, depending on which shrunk lattice we are considering and, regardless of the color, they can be of type $X$ or type $Z$. We  denote these string operators by $S^{C \sigma}_{\mu}$ where $C$ is a color, $\sigma$ is $ Z $ or $ X $ and $\mu$ is a label of the homology class.

In general, one has \cite{distillation}:

\begin{equation} \label{indep color}
S^{R \sigma}_{\mu}S^{G \sigma}_{\mu}S^{B \sigma}_{\mu} \sim 1 \,.
\end{equation}  

The equation (\ref{indep color}) shows that one has only two independent colors.

In order to form a base of Pauli operators acting on the qubits, we consider strings which are not homologically trivial, as it is done in the surface codes. Two strings of the same type always commute. Two strings of the same color always have an even number of qubits in common, so they also commute. In this way, the only way to have strings that anti-commute is if we have strings with different types, with different colors and with an odd number of intersections.

\section{New Families of Color Codes}

In the case of trivalent and $3$-colorable tessellations by regular polygons in the Euclidean plane, we are restricted to only three cases, one regular and two semirregular: $\{6, 6, 6\}$, $\{4, 8, 8\}$ and $\{4, 6, 12\}$. 

The notation $\{p,q\}$ indicates a regular polygon with $p$ sides, that tessellates the hyperbolic plane, such that in each vertex of the tessellation there are $q$ other polygons. And, by $\{p,q,r\}$ we mean a tessellation of the plane where in each vertex one has the meting of regular polygons with $p$, $q$ and $r$ sides, respectively, in this order.

By increasing the genus of the surface, and hence the number of coded qubits, there are no polygons which are flat models of these surfaces in the Euclidean plane, but there exist in the hyperbolic plane.

Hyperbolic polygons which are flat models for surfaces of genus $g \geq 2$ are called fundamental polygons \cite{beardon2012geometry}.

The advantage in this case is that there are infinite possibilities of trivalent tessellations of the hyperbolic plane. Even excluding those that are not $3$-colorable, we still have infinite possibilities.

From \cite{magnus} one has that $\{p,3\}$ is a trivalent regular tessellation of the hyperbolic plane if, and only if, it satisfies $\frac{1}{p} +\frac{1}{3} < \frac{1}{2}$. 

\begin{figure}[!h]
	\centering
	\includegraphics[scale=0.45]{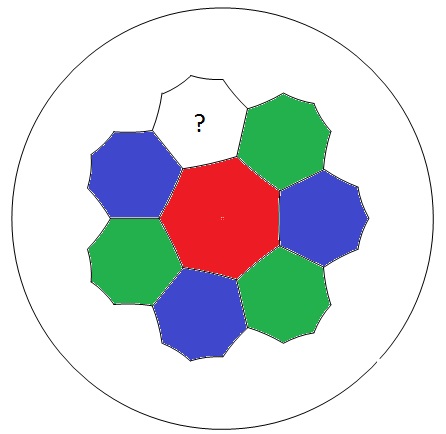} 
	\caption{A hyperbolic plane tesselation by a \{7,3\} polygon. Even though it is a trivalent tessellation, it is not 3-colorable.}	\label{7-3}
\end{figure}

As already quoted, many of these tessellations, though trivalent, are not $3$-colorable. As an instance, $\{7, 3\}$ (see figure the (\ref{7-3})), or anyone that it has a polygon with an odd number of sides, are not $3$-colorable.

Given a fundamental polygon $P^{\prime}$ that generates a $g$-torus, we consider a tessellation of it by regular polygons $P$. We must determine how many polygons $ P $ are needed to exactly cover the area of $P^\prime$, that is, we must determine the solutions of:

\begin{equation} \label{area}
\mu(P^\prime)=n_f \mu(P)
\end{equation}

\noindent where $\mu(X)$ denotes the area of $X$ and $n_f$ is the number of faces of the tessellation $\{p,q\}$, with $p>6$. Then, one has

\begin{equation} \label{faces}
n_f=\frac{4q(g-1)}{pq-2p-2q} \,.
\end{equation}

Since we will use only trivalent tesselations, we set $ q = 3 $. In particular, for the double torus we have $g=2$, then

\begin{equation}
n_f=\frac{12}{p-6}; p>6 \,,
\end{equation}

which generates the possibilities described in the Table~1. 

\begin{table}[!h]
	\centering
	\label{bitoro}
	\begin{tabular}{|l|l|}
		\hline
		$p$  & $n_f$ \\ \hline
		8  & 6 \\ \hline
		10 & 3 \\ \hline
		12 & 2 \\ \hline
		18 & 1 \\ \hline
	\end{tabular}\vspace{4pt}
	\caption{Number of faces of the tessellation according to the number of sides of the polygon which gives rise to a surface with genus $2$}
\end{table}

Polygons with $12$ and $18$ sides are not useful, since they generate tessellations with less than $3$ faces. Let us analyse the remaining cases.

The distance of a stabilizer code is the weight of the Pauli Operator with the lower weight which preserves the code subspace, and acts non trivially on it. In terms of color codes, it is the smallest number of qubits in the support of a homologically non-trivial cycle of the tessellation, looking to the shrunk lattice.

Using formulas from hyperbolic trigonometry, we obtain a lower bound for the minimum distance of the codes generated by the given tessellations.

Since the minimum distance of the code depends of a "path" on the edges and faces of the tessellation, we have that the hyperbolic length of a path on edges and faces of the tessellation connecting opposite sides of a polygon is greater than the hyperbolic length of a geodesic connecting these same sides.

Thus, given a fundamental polygon of $\{4g, 4g\}$, considering a pairing by opposite sides, the hyperbolic distance $ d_h $ between paired sides can be calculated by

\begin{equation} \label{dh}
d_h = 2 arccosh \left[ \frac{\cos(\pi/4g)}{\sin(\pi/4g)} \right] \,,
\end{equation}

\noindent as we can see in \cite{beardon2012geometry}.

Let us denote the hyperbolic length of each edge of a tessellation $\{p, q\}$ by $l_{p,q}$. Thus:

\begin{equation} \label{lado}
l_{p,q} = arccosh \left[ \frac{\cos^2(\pi/q) +\cos(2\pi/p)}{\sin^2(\pi/q)} \right] \,.
\end{equation}

Also, the greatest distance between two points of a hyperbolic polygon is the distance between two opposite vertices (if any), which may be bounded by twice the lenght of the radius of the hyperbolic circumscribed circle to the polygon.

Given a regular polygon of $\{p, q\}$, the diameter of its circumscribed circle, denoted by $D_{p,q}$, may be calculated by:

\begin{equation} \label{diametro}
D_{p,q} = 2 arccosh \left[ \frac{\cos(\pi/p) \cos(\pi/q)}{\sin(\pi/p) \,. \sin(\pi/q)} \right]  \,.
\end{equation}

In this way we may calculate an upper bound for an edge of the shrunk lattice, adding the length of an edge of the tessellation with the diameter of the circumscribed circle to a polygon of that tessellation. Let us denote this value by $AR_{p,q}$. Thus,

\begin{equation} 
AR_{p,q} = l_{p,q}+D_{p,q} \,.
\end{equation}

This measure allows us to calculate a lower bound for the number of edges of the reduced network contained in a non-trivial homology cycle belonging to such a lattice, and thus a bound for the minimum distance of the code. That is,

\begin{equation}
n_a > \frac{d_h}{AR_{p,q}} \,.
\end{equation}

It follows that the minimum distance of the code is determined by

\begin{equation} \label{distancia}
d= 2 \lceil \frac{d_h}{AR_{p,q}} \rceil \,.
\end{equation}

From the tessellations in the Table~1, let us consider polygons of $\{8,3\}$ and $\{10,3 \}$. To the tessellation $\{8,3\}$ one has, by (\ref{lado}),

\[
l_{8,3} = arccosh \left[ \frac{\cos^2(\pi/3) +\cos(2\pi/8)}{\sin^2(\pi/3)} \right]
\approx 0,7270398 \,,
\]

and, by (\ref{diametro}),

\[
D_{8,3} = 2 arccosh \left[ \frac{\cos(\pi/8) \cos(\pi/3)}{\sin(\pi/8) \sin(\pi/3)} \right] \approx 1,721412 \,,
\]

giving that 
\[
AR_{8,3} \approx 2,448452 \,.
\]

Therefore, we may calculate the minimum distance of the code using (\ref{distancia}):

\[
d = 2 \lceil \frac{d_h}{AR_{8,3}} \rceil = 4 \,.
\]

Now, to the tessellation $\{10, 3\}$ one has:

\[
l_{10,3} \approx 0,87917928 
\] 

and

\[
D_{10,3} \approx 2,354664 \Rightarrow AR_{10,3} \approx 3,23384 \,,
\]

which implies that 

\[
d= 2 \lceil \frac{d_h}{AR_{10,3}} \rceil = 2 \,.
\]

Thus, we have generated codes with parameters $ [[16, 8, 4]] $ and $ [[10, 8, 2]]$, respectively, which are different parameters in relation to the families previously presented in \cite{Bombin2006Topological}, \cite{Kitaev} and \cite{distillation}, among others.

\begin{figure}[!h]
	\centering
	\includegraphics[scale=0.45]{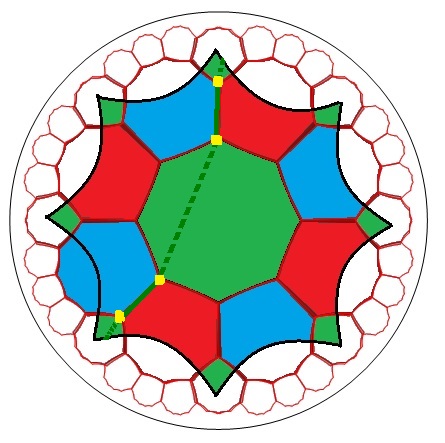}
	\caption{A $\{8, 8\}$ polygon tessellated by a $\{8,3\}$ giving rise to a code with parameters $[[16, 8, 4]]$. Explicitly a green string of nontrivial homology with its support, which are the four qubits marked in yellow.}
	\label{8-3blue}
\end{figure}

With calculations and arguments analogous to those made for the double torus, we will generate codes on compact surfaces of genus greater than $2$, that is, some $g$-torus, and analyze their parameters.

As it was mentioned, for each $g$-torus considered, we will use a fundamental polygon of $\{4g,4g\}$ and the pairing of edges identifying opposite sides.

If we set the minimum distance of the code that we want to generate, for instance $ d = 4 $, we may generate the code family using the technique presented here.

We observe that the given distance $ d = 4 $ is determined by the code generated by a tessellation of a compact two-dimensional surface such that there are exactly $6$ faces of the tessellation. Thus, by (\ref{faces}), using $ n_f = 6 $ and $ q = 3 $ one has:

\begin{eqnarray}
n_f  &=&  \frac{4q(g-1)}{pq-2p-2q} \nonumber\\
6    &=& \frac{12(g-1)}{p-6}  \nonumber \\
p-2g &=& 4 \nonumber\\
p &=& 4+2g 
\end{eqnarray}

\noindent which shows that for any given genus, with $ g \geq 2 $, we are able to create a code with the given minimum distance $ d = 4 $. Thus, we obtain the code family $ [[8 + 4g, 4g, 4] $.

This family of codes has the remarkable property that 

\[
\lim_{n\rightarrow \infty} \frac{k}{n} =1 \,.
\]

In the following tables we present examples of color codes generated by this method, on compact surfaces of genera $3$ to $9$. In each table the value of $d_h$ was giving by the formula (\ref{dh}).

\begin{table}[h]
	\centering
	\label{3toro}
	\begin{tabular}{|l|l|l|l|l|}
		\hline
		$\{p,q\}$  & $n_f$ & $AR_{p,q}$ & $d_h/AR_{p,q}$ & $[[n,k,d]]$   \\ \hline
		$\{8,3\}$  & 12    & 2,44845    & 1,23176          & $[[32,12,4]]$ \\ \hline
		$\{10,3\}$ & 6     & 3,23384    & 1,62687          & $[[20,12,4]]$ \\ \hline
		$\{14,3\}$ & 3     & 4,15197    & 0,95937          & $[[14,12,2]]$ \\ \hline
	\end{tabular}\vspace{4pt}
	\caption{Parameters of the color codes generated in the $3$-tori according to each tessellation, using $ d_h \approx 3,9833 $}
\end{table}

\begin{table}[h]
	\centering
	\label{4toro}
	\begin{tabular}{|l|l|l|l|l|}
		\hline
		$\{p,q\}$  & $n_f$ & $AR_{p,q}$ & $d_h/AR_{p,q}$ & $[[n,k,d]]$   \\ \hline
		$\{8,3\}$  & 18    & 2,44845    & 1,87710          & $[[48,16,4]]$ \\ \hline
		$\{10,3\}$ & 9     & 3,23384    & 1,42122          & $[[30,16,4]]$ \\ \hline
		$\{12,3\}$ & 6     & 3,75563    & 1,22376          & $[[24,16,4]]$ \\ \hline
		$\{18,3\}$ & 3     & 4,74604    & 0,96838          & $[[18,16,2]]$ \\ \hline
	\end{tabular}\vspace{4pt}
	\caption{Parameters of the color codes generated in the $4$-tori according to each tessellation, using $d_h \approx 4,596$}
\end{table}

\begin{table}[]
	\centering
	\label{5toro}
	\begin{tabular}{|l|l|l|l|l|}
		\hline
		$\{p,q\}$  & $n_f$ & $AR_{p,q}$ & $d_h/AR_{p,q}$ & $[[n,k,d]]$   \\ \hline
		$\{8,3\}$  & 24    & 2,44845    & 2,06624        & $[[64,20,6]]$ \\ \hline
		$\{10,3\}$ & 12    & 3,23384    & 1,56442        & $[[40,20,4]]$ \\ \hline
		$\{14,3\}$ & 6     & 4,15197    & 1,21848        & $[[28,20,4]]$ \\ \hline
		$\{22,3\}$ & 3     & 5,19193    & 0,97441        & $[[22,20,2]]$ \\ \hline
	\end{tabular}\vspace{4pt}
	\caption{Parameters of the color codes generated in the $5$-tori according to each tessellation, using $d_h \approx 5,0591$}
\end{table}

\begin{table}[]
	\centering
	\label{6toro}
	\begin{tabular}{|l|l|l|l|l|}
		\hline
		$\{p,q\}$  & $n_f$ & $AR_{p,q}$ & $d_h/AR_{p,q}$ & $[[n,k,d]]$   \\ \hline
		$\{8,3\}$  & 30    & 2,44845    & 2,21885        & $[[80,24,6]]$ \\ \hline
		$\{10,3\}$ & 15    & 3,23384    & 1,67997        & $[[50,24,4]]$ \\ \hline
		$\{16,3\}$ & 6     & 4,47385    & 1,21433        & $[[32,24,4]]$ \\ \hline
		$\{26,3\}$ & 3     & 5,55117    & 0,97866        & $[[26,24,2]]$ \\ \hline
	\end{tabular}\vspace{4pt}
	\caption{Parameters of the color codes generated in the $6$-torus according to each tessellation, using $d_h \approx 5,43275$}
\end{table}

\begin{table}[]
	\centering
	\label{7toro}
	\begin{tabular}{|l|l|l|l|l|}
		\hline
		$\{p,q\}$  & $n_f$ & $AR_{p,q}$ & $d_h/AR_{p,q}$ & $[[n,k,d]]$   \\ \hline
		$\{8,3\}$  & 36    & 2,44845    & 2,34687        & $[[96,28,6]]$ \\ \hline
		$\{10,3\}$ & 18    & 3,23384    & 1,77697        & $[[60,28,4]]$ \\ \hline
		$\{12,3\}$ & 12    & 3,75563    & 1,53009        & $[[48,28,4]]$ \\ \hline
		$\{14,3\}$ & 9     & 4,15197    & 1,38403        & $[[42,28,4]]$ \\ \hline
		$\{18,3\}$ & 6     & 4,74604    & 1,21079        & $[[36,28,4]]$ \\ \hline
		$\{30,3\}$ & 3     & 5,85296    & 0,98180        & $[[30,28,2]]$ \\ \hline
	\end{tabular}\vspace{4pt}
	\caption{Parameters of the color codes generated in the $7$-torus according to each tessellation, using $d_h \approx 5,7464$}
\end{table}

\begin{table}[]
	\centering
	\label{8toro}
	\begin{tabular}{|l|l|l|l|l|}
		\hline
		$\{p,q\}$  & $n_f$ & $AR_{p,q}$ & $d_h/AR_{p,q}$  & $[[n,k,d]]$   \\ \hline
		$\{8,3\}$  & 42    & 2,44845    & 2,45747        &$[[112,32,6]]$ \\ \hline
		$\{10,3\}$ & 21    & 3,23384    & 1,86063        & $[[70,32,4]]$ \\ \hline
		$\{20,3\}$ & 6     & 4,98250    & 1,20763        & $[[40,32,4]]$ \\ \hline
		$\{34,3\}$ & 3     & 6,11364    & 0,984193       & $[[34,32,2]]$ \\ \hline
	\end{tabular}\vspace{4pt}
	\caption{Parameters of the color codes generated in the $8$-torus according to each tessellation, using $d_h \approx 6,01699$}
\end{table}

\begin{table}[]
	\centering
	\label{9toro}
	\begin{tabular}{|l|l|l|l|l|}
		\hline
		$\{p,q\}$  & $n_f$ & $AR_{p,q}$ & $d_h/AR_{p,q}$ & $[[n,k,d]]$   \\ \hline
		$\{8,3\}$  & 48    & 2,44845    & 2,55465        &$[[128,36,6]]$ \\ \hline
		$\{10,3\}$ & 24    & 3,23384    & 1,93422        & $[[80,36,4]]$ \\ \hline
		$\{14,3\}$ & 12    & 4,15197    & 1,50650        & $[[56,36,4]]$ \\ \hline
		$\{22,3\}$ & 6     & 5,19193    & 1,20474        & $[[44,36,4]]$ \\ \hline
		$\{38,3\}$ & 3     & 6,34331    & 0,98607        & $[[38,36,2]]$ \\ \hline
	\end{tabular}\vspace{4pt}
	\caption{Parameters of the color codes generated in the $9$-torus according to each tessellation, using $d_h \approx 6,254948$}
\end{table}

We can see in the tables that all these codes have minimum distance $d \geq 2$. In the cited examples, all the codes satisfying the quantum Hamming and Singleton bound. 

About the quantum Singleton bound, which can be written according to the parameters of the code in the form $ n - k \geq 2d-2 $, we have that in all the tables presented, the last row of the table has a code that saturates such bound, that is, satisfies the bound with equality. This is a fact that also calls attention, because it is always the code with the highest encoding rate in each surface, although it is the one with the smallest minimum distance.

Some of the codes presented, such as codes $[[16,8,4]]$, $[[20, 12, 4]]$, $[[24, 16, 4]]$ and $[[28, 20, 4]]$, are better than best-in-category in Grassl's coding tables \cite{markus}.

\end{document}